\documentclass[letterpaper,twocolumn]{article}
\usepackage[raggedright]{titlesec}
\usepackage{amsmath}
\usepackage{amsfonts}
\usepackage{color,soul}
\usepackage{xcolor}
\usepackage{colortbl}

\usepackage{pbox}
\usepackage{caption}
\usepackage{subcaption}
\usepackage{array,multirow,graphicx}
\usepackage{floatrow}
\usepackage{balance}

\usepackage{graphicx}
\usepackage{amsmath}
\usepackage{amsfonts}
\usepackage{ amssymb }
\usepackage{booktabs}
\usepackage{tabularx}
\usepackage{multirow}
[]
\date{}

\titlespacing{\section}{0pt}{*2.0}{*2.0}
\titlespacing{\subsection}{0pt}{*1.8}{*1.8}

\graphicspath{ {figs/} }

\begin{document}

\title{\Large\textbf{Protein Mutation Stability Ternary Classification using Neural Networks and Rigidity Analysis}\normalsize}

\author{
    Richard Olney,\textsuperscript{1} Aaron Tuor,\textsuperscript{2} 
   % Max Shelbourne,\textsuperscript{1}\\
    Filip Jagodzinski\textsuperscript{1} and Brian Hutchinson\textsuperscript{1,2}\\
    \textsuperscript{1}Western Washington University, Bellingham, WA, USA\\
    \textsuperscript{2}Pacific Northwest National Laboratory, Seattle, WA, USA\\
    \{filip.jagodzinski, brian.hutchinson\}@wwu.edu\\
}

\maketitle 

\thispagestyle{empty}

\begin{center}
\large\textbf{Abstract}
\end{center}

\vspace{2mm} Discerning how a mutation affects the stability of a
protein is central to the study of a wide range of diseases. Machine
learning and statistical analysis techniques can inform how to
allocate limited resources to the considerable time and cost
associated with wet lab mutagenesis experiments. In this work we
explore the effectiveness of using a neural network classifier to predict the change in the stability of
a protein due to a mutation.
Assessing the accuracy of our approach is dependent on the use of
experimental data about the effects of mutations performed \textit{in
  vitro}. Because the experimental data is prone to discrepancies when
similar experiments have been performed by multiple laboratories, the
use of the data near the juncture of stabilizing and destabilizing
mutations is questionable. We address this later problem via a
systematic approach in which we explore the use of a three-way
classification scheme with stabilizing, destabilizing, and inconclusive
labels. For a systematic search of potential classification cutoff
values our classifier achieved 68 percent accuracy on ternary
classification for cutoff values of -0.6 and 0.7 with a low rate of
classifying stabilizing as destabilizing and vice versa.

\medskip
\noindent
%\textbf{keywords:} This part is optional. The word \textbf{keywords} is flushed to the left (no indentation) with 10-point bold font. Include up to 6 keywords. %One space line below the abstract.

\section{Introduction}
\noindent 
Performing an amino acid substitution in a protein may induce a
structural change that can have wide ranging effects on the protein's
function.  Discovering which mutations are destabilizing and which are
stabilizing provides insights into many types of disorders, such as
sickle cell anemia \cite{serjeant1992sickle} and some types of cancer
\cite{figueroa2010leukemic}, and is important for understanding
communicable and highly mutable diseases (e.g. HIV
\cite{he1995human}, influenza \cite{virus2009emergence}).

\textit{In vitro} experiments are necessary to determine how a
mutation affects a protein's function. However, these experimental
efforts come at considerable time and cost, as a single mutagenesis
experiment followed by X-ray crystallography work may require weeks of wet lab
work. Moreover, because each residue in a protein can in principle be
one of 20 naturally occurring amino acids, the set of all possible
mutations is vast, so computational tools for screening likely
candidates for investigation in a wet lab setting are desired.

We explore the use of a neural network classifier for automatic
inference of the effects of mutations. 
The ground truth, obtained from wet lab experiments recorded in the Protherm database \cite{kumar2006protherm}, 
 is in the form of change of the Gibbs Free Energy ($\Delta\Delta
G$) indicating whether a mutation is destabilizing (negative
$\Delta\Delta G$) or stabilizing (positive $\Delta\Delta G$). 
Typical approaches either predict the $\Delta\Delta G$ value given a specified mutation (regression) \cite{capriotti2005, farhoodi2017predicting}, 
or predict whether a mutation is stabilizing or destabilizing (binary classification) \cite{capriotti2005}.  

Here we deal with ternary classification in which a third
``inconclusive'' class is introduced. That class is important because
all available $\Delta\Delta G$ data is from wet-lab
work, and as with any physical experiment, there is the chance of some
inherent error. The use of a $\Delta\Delta G$ value close to $0$ might
cause a classifier to mis-classify a stabilizing mutation as
destabilizing or vice versa, if indeed the reported true label is
erroneous. Mislabeled data is detrimental to training a  model, so we systematically performed many computational experiments, 
testing the range of indeterminate values to find an optimal
inconclusive range for $\Delta\Delta G$.

We trained deep neural network classification models across a
systematic search of the $\Delta\Delta G$ cutoff space. Using the
results of these experiments we generated confusion matrices in
order to assess the utility and classification performance for each cutoff range. We
found several interesting trends and potential cutoff ranges, which we
present here via case studies.

\section{Related Work and Motivation}
\label{motivation}
The use of experimental stability data ($\Delta\Delta G$) is prevalent
in research that aims to offer computational techniques for assessing
the effects of mutations~\cite{kortemme2002,chen2013, kang2008}. An
often-cited source is the ProTherm database~\cite{kumar2006protherm}. It
provides information about the proteins, mutations performed, wet lab
conditions, and stability measurements for 25,820 mutation experiments
reported on in the literature. Of those ProTherm entries for
which stability data is provided, the $\Delta\Delta G$ values range
from about -10 kCal/mol (indicating a strongly destabilizing
mutation), to approximately +10 kCal/mol (strongly stabilizing). The
single inflection value of zero $\Delta\Delta G$ designates that point
on the real number line where the effect of a mutation changes from
stabilizing to destabilizing.

\begin{table}
  \caption{Distribution of $\Delta\Delta G$ values among ProTherm entries for which stabilizing information is available}
  \label{proThermEntries}
  \centering
  \begin{tabular}{|c|c|c|}
    \hline
    \cellcolor{lightgray}$\Delta\Delta G$ lower val &\cellcolor{lightgray} $\Delta\Delta G$ upper value &\cellcolor{lightgray} Num Entries\\
    \hline
    -10 & 10 & 4184\\ 
    \hline
    -1 & 1 & 2157\\
    \hline
    -0.5 & 0.5 & 1364\\
    \hline
    -0.1 & 0.1 & 390\\
    \hline
  \end{tabular}
\end{table}

In Table~\ref{proThermEntries} we show the count of entries in
ProTherm for three separate ranges of $\Delta\Delta G$ values. Of the
4,184 entries with $\Delta\Delta G$ ranges between -10 and 10
kCal/mol, 1,364 of them are in the range $[-0.5 , +0.5]$. Thus, a
large portion of ProTherm entries are for values where experimental
errors or instrument discrepancies might mean that a recorded
stabilizing mutation is indeed destabilizing, and vice versa. It is for
this reason that experimental data for $\Delta\Delta G$ values in the
range $[-0.5 , +0.5]$ is often not used.

In addition, there are a number of entries in the ProTherm database
where identical experiments performed by different labs have recorded
opposite (stabilizing versus destabilizing) results. Two examples :

\begin{itemize}
\item Cold shock Protein, ProTherm Entries 21797 and 21839,
  $\Delta\Delta G$ = -0.05 and +0.7, respectively
\item Myoglobin Sperm Whale, ProTherm Entries 2092 and 2814,
  $\Delta\Delta G$ = -0.9 and + 0.1, respectively
\end{itemize}

The use of $\Delta\Delta G$ data, therefore, as values for assessing  and training a predictive model, must be done with care. For this reason, we report the predictive power of our machine learning model in the context of a systematic approach of varying the $\Delta\Delta G$ values designating boundaries between three classification labels.

\section{Methodology}
Here we summarize rigidity analysis and describe
how we generate features and labels for training our neural network classifier machine learning model, and the experiment setup for
evaluating the model.

\subsection{Rigidity Analysis}
\begin{figure}
  \centering
\begin{subfigure}[b]{0.47\textwidth}
        \includegraphics[width=\textwidth]{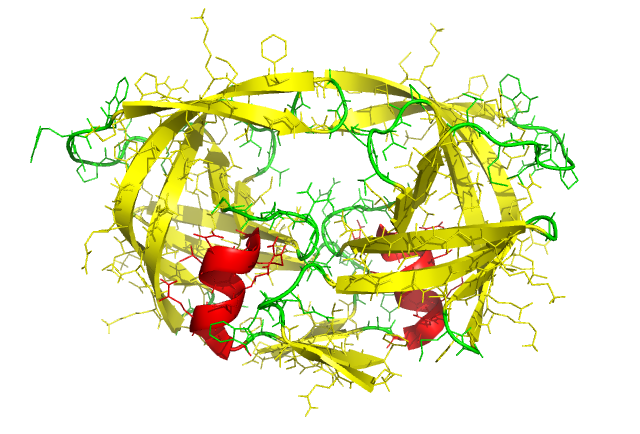}
        \caption{Cartoon}
    \end{subfigure}
    \begin{subfigure}[b]{0.42\textwidth}
        \includegraphics[width=\textwidth]{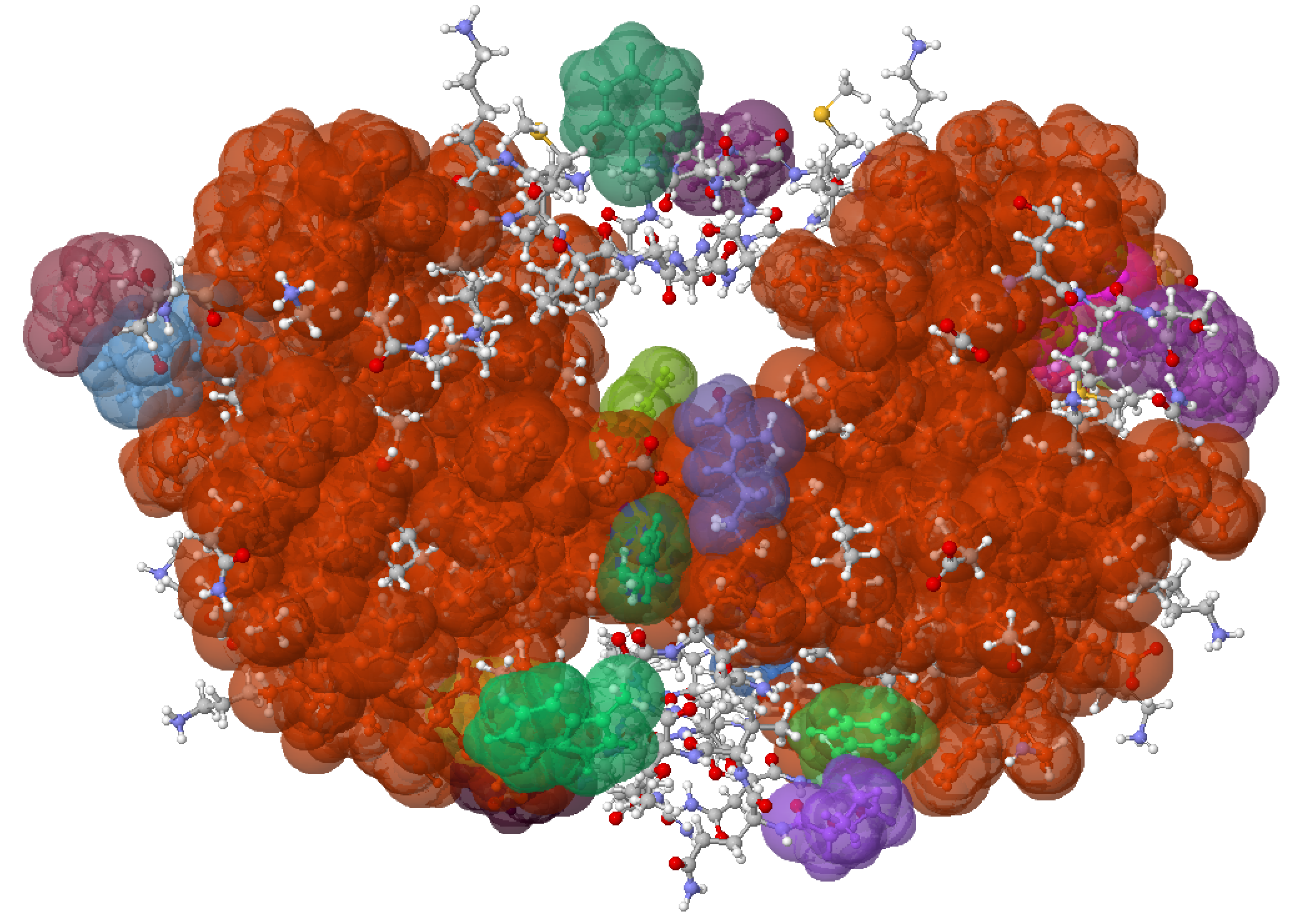}
        \caption{Rigidity Analysis}
    \end{subfigure}
   \caption{Rigidity analysis (PDB 1HVR) identifies rigid
     clusters. Orange is the largest cluster with 1,371 atoms.}
   \label{rigAnalysis}
\end{figure}

To help reason about the effects of mutations, we take an approach
that relies on a fast algorithm for assessing the
rigidity of a protein~\cite{foxlibrary2012,jacobs2001}. In rigidity
analysis, atoms and their chemical interactions are used to construct
a mechanical model, a graph is constructed from the model, and pebble
game algorithms~\cite{jacobs1997} are used to analyze the
rigidity of the associated graph. The results are used to infer the
rigid regions of the protein (Figure~\ref{rigAnalysis}). We rely on the KINARI
rigidity software for performing rigidity
analysis~\cite{foxlibrary2012}.

\subsubsection{Mutants, Rigidity Distance}
To generate \textit{in silico} mutant structures corresponding to the
mutation data in ProTherm, we used our ProMuteHT~\cite{andersson2017}
software. In this study, we rely on the rigidity analysis results of
the wild type (non-mutated protein), and a mutant, to assess the
effects of a mutation. In our previous work~\cite{andersson2016, farhoodi2017predicting}, we
used an $RD_{WT\rightarrow mutant}$ rigidity distance metric to
quantitatively assess the impact of mutating a residue to one of the
other 19 naturally occurring amino acids:

\begin{center}
\vspace{0.2cm}
$RD_{WT \rightarrow mutant}
: \sum_{i=1}^{i=LRC} i \times [WT_i - Mut_i]$
\vspace{0.2cm}
\end{center}

\noindent
where $WT$ refers to Wild Type, $Mut$ refers to mutant,
and $LRC$ is the size of the Largest Rigid Cluster (in
atoms). Each term of the summation $RD_{WT\rightarrow mutant}$ metric calculates
the difference in the count of a specific cluster size, $i$, of the
wild type and mutant, and weighs that difference by $i$.

\subsection{Wet Lab Mutation Data -- $\Delta\Delta G$}
\begin{figure}
	\center
	\includegraphics[scale=0.37]{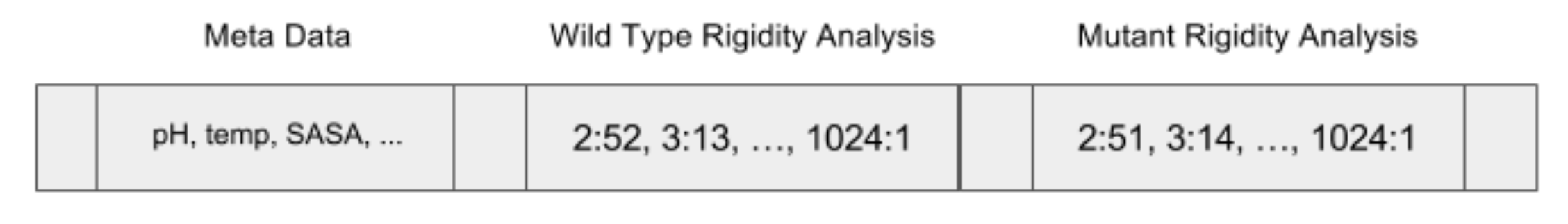}
    \caption{The form of a feature vector as input to the DNN. It
      consists of experiment meta data, such as Solvent Accessible
      Surface Area (SASA), pH, and temperature, concatenated with the
      rigid cluster frequencies of both wild type and mutant
      proteins.}
    \label{features}
\end{figure}

Labels ($\Delta\Delta G$) and metadata (pH, temperature, etc.) of
mutations were retrieved from the ProTherm \cite{kumar2006protherm} database of
mutation experiments. The rigidity features for each mutant and wild
type were generated by rigidity analysis using the KINARI software. A
total of $2,072$ data points from ProTherm meet the criteria for our
experimental setup (i.e., single chain proteins, single mutations, any
value of $\Delta\Delta G$).  The input to our model is shown in
Figure~\ref{features}.  The data set of 2,072 proteins is split into a
training set of 1,438 proteins for fitting a classifier, a development
set of 324 proteins for finding the best neural network configuration,
and a test set of 310 proteins to test generalization error.

\subsection{Deep Neural Network Classifier}
A deep neural network (DNN) classifier is a parameterized function mapping a real valued vector to a probability distribution over a set of classes. 
%The parameters are typically fit using first order optimization methods, with gradients computed efficiently via backpropagation \cite{rumelhart86learning}. 
%The classic neural network
%contains a single hidden layer (see Eqn. \ref{eqn:layer}) and an output activation
%layer (Eqns \ref{eqn:softmax} and \ref{eqn:output}) which converts the function's representation
%into a probability distribution across the classes. 
%A deep neural
%network  differs from a classic neural network in that it may contain
%an arbitrary number of hidden layers. 
We model the probability distribution over classes
 of mutation as stabilizing, destabilizing, or inconclusive, as a  function of the rigidity analyses and experimental conditions,
using a DNN with $L$ hidden layers,  $\textbf{h}_{(1)}, \textbf{h}_{(2)}, \dots, \textbf{h}_{(L)}$. 
This neural network classifier takes as input a feature vector ${\bf x}$ (Fig. \ref{features})  which we alternatively denote as
as ${\bf h}_{(0)}$. The classifier
 outputs a probability vector ${\bf p} \in \mathbb{R}^3$, the elements of which are
 calculated as:
\begin{eqnarray}
    \textbf{p}_{k} & = & \frac{\texttt{exp}({{\bf o}_k})}{\sum_{j=1}^3 \texttt{exp}({{\bf o}_{j}})} \mbox{, where} \label{eqn:softmax}\\
    {\bf o} & = &  {\bf U} {\bf h}_{(L)} + {\bf a} \mbox{ and } \label{eqn:output}\\
    {\bf h}_{(\ell)} & = & f({\bf W}_{(\ell)} {\bf h}_{(\ell-1)} + {\bf b}_{(\ell)}).\label{eqn:layer}
\end{eqnarray}

\noindent
where hidden activation function $f$ is one of three non-linear
functions operating elementwise on matrices; the hyperbolic tangent function ($\tanh$), the logistic
sigmoid function, or the rectified linear unit function
($\mbox{ReLU}$). The trainable parameters are
the $L$ hidden weight matrices ($\bf{W}$), $L$ bias vectors
($\bf{b}$), and the output layer weights and bias $\bf{U}$ and
$\bf{a}$. 

All model parameters were trained with the Adam optimization algorithm
\cite{kingma2014adam}, a variant of stochastic gradient descent.  The
training loss is the cross-entropy between the true distribution as
determined by inconclusive bounds and the DNN's predicted distribution.

The DNN hyper-parameters are model choices which cannot be learned via the training data through gradient descent. They are instead selected 
by evaluating models on the held out development set which is distinct from the training data and the testing set. The model choices we select in this fashion are
the number of hidden layers, the size of each hidden layer (dimensions of the weight matrices $\textbf{W}$), the hidden activation function, and finally, the mini-batch size and learning rate used in stochastic gradient descent optimization.

We developed and trained our model architecture using the 
 Tensorflow \cite{tensorflow2015-whitepaper} Python library.
% for its support of back propagation and and its implementation of Nvidia's CUDA GPU parallelization. 
 Due to the small data set and GPU acceleration for
computation, it takes under a minute to train a typical model.  
\subsection{Class Labels}
As already mentioned, the
$\Delta\Delta G$ values in ProTherm -- especially those near zero --
must be used with caution. To help determine which range of
$\Delta\Delta G$ values should delimit stabilizing, destabilizing, and
inconclusive mutations, we employed a principled approach by training
models across a systematic set of different inconclusive ranges to
train the best predictive model.

Class labels are represented as probability distributions over the
three classes, i.e. real valued vectors in $\mathbb{R}^3$ that contain
non-negative values and sum to one.  A label for $\Delta\Delta G$
classification has one element as 1 and the other elements are
zero. So,
$[1 \;\; 0 \;\; 0]^T$
%$\begin{bmatrix}
%1\\
%0\\
%0
%\end{bmatrix}$ 
corresponds to a $\Delta\Delta G$ score which is negative and outside
the range of indeterminacy (a destabilizing mutation),
%$\begin{bmatrix} 0\\ 1\\ 0
%\end{bmatrix}$ 
$[0 \;\; 1 \;\; 0]^T$ corresponds to an inconclusive $\Delta\Delta G$
score inside the range of indeterminacy, and $[0 \;\; 0 \;\; 1]^T$
%$\begin{bmatrix} 0\\ 0\\ 1
%\end{bmatrix}$ 
corresponds to a $\Delta\Delta G$ score which is positive and outside
the range of indeterminacy (a stabilizing mutation).
%
%For a given vector if the first position is a 1 the vector is
%predicting the mutation as destabilizing, if the middle position is a
%one the model is predicting the mutation as inconclusive, and if the
%final third position is a 1 the model is predicting the mutation to be
%stabilizing. 
To make a prediction from our model's predicted class distribution,
${\bf p}$, we pick the most probable index.
%Our model is constrained such that it always produces a
%sparse representation, meaning there is only ever a single 1 in the
%vector and the rest of the entries are 0.
 
Our ultimate goal is to find a pair of $\Delta\Delta G$ values for
which a model can be trained to correctly predict the true labels that
those cutoffs would create. For example if the $\Delta\Delta G$ value
of an experiment is reported to be -0.8, and our model's $\Delta\Delta
G$ cutoffs were -0.6 and 0.7,
%Figure \ref{cmcutoff-0607} 
the true label for that mutation would be destabilizing and a correct
prediction from our model would also be destabilizing. For our
best model, predictions should match true labels as closely as
possible.

\subsection{Experimental Setup}
In order to assess our model's effectiveness at classification for
different inconclusive bounds, we trained 100 DNN models with
random hyper-parameter configurations (the same configurations were used for all cutoff ranges).
%(constrained to be ReLu if number of hidden layers was
%greater than four)
%These 100 different hyper-parameter configurations were cached and reused for each inconclusive cutoff range. 
We normalized $\Delta\Delta
G$ by dividing all values by 10, % so that their magnitudes would roughly
%range between -1 and 1. 
and executed a triangular grid search of cutoff ranges equivalent to -2.0 to 2.0,
stepping by 0.1, in unnormalized $\Delta\Delta G$, for a total of 820 cutoff ranges. 
All 100 hyper-parameter configurations were assessed for
each range, for a total of 8,200 configurations. 

\section{Confusion Matrices}
\begin{figure}
  \centering
    \includegraphics[width=0.75\textwidth]{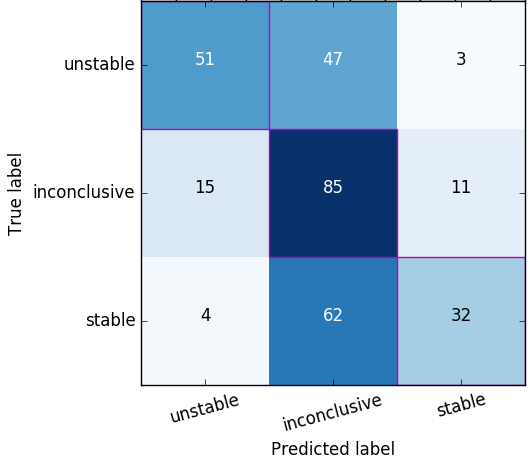}
    \caption{Test set confusion matrix for cutoffs -0.5, 0.5.}
    \label{cmcutoff0505}
  \end{figure}

For each of the 820 cutoff  ranges, we identified the DNN model
which achieved the best development set accuracy, and generated a confusion matrix
for those model's predictions on the held out test set. Confusion
matrices are a method of visualizing the performance of the
classification algorithm. They contain the same classes on the
vertical and horizontal axis, with the vertical axis indicating true
labels for each class and the horizontal axis indicating the model's
class predictions. Figure \ref{cmcutoff0505} is the confusion matrix
generated by the classic heuristic for inconclusive $\Delta\Delta G$
of -0.5 to 0.5. The darker the color the more predictions fall into
that intersection of true label and predicted label. A perfect
classification model would have predictions only in the top left, center, and
bottom right squares.

In addition to the standard metrics of a model's accuracy in
predicting the correct class, the confusion matrices offer insights in
cases when a model is mis-classified. They allow us to assess Type I
and Type II errors, false positive and false negative classifications,
and also permit seeing how those incorrect classifications are being
classified.  This additional information enables assessing whether a
particular mis-classification is more detrimental than another. For
instance it may be better if a model is less accurate overall, but
predicts very few unstable mutations as stabilizing and vice versa,
but has a slightly higher than ideal tendency to label mutations as
inconclusive.

\begin{table*}
	\begin{tabularx}{\linewidth}{l  cc  c cc c  cc  c c c c c c ccccc }
	&&\multicolumn{5}{l}{\textbf{Hyper-parameters}}&&&&\multicolumn{2}{l}{\textbf{Range}}&&&\multicolumn{4}{l}{\textbf{Metrics}}\\
	\toprule
		{\bf Model}&&{\bf mb}&{\bf lr}&{\bf hs} &{\bf nl}&{\bf ha}&&&&{\bf L}&{\bf U}&&&{\bf loss}&{\bf acc}&{\bf macc}&{\bf ratio}\\
		\midrule
                Figure \ref{cmcutoff0505}&&64 &0.01 &689 & 1&sigmoid&&&&0.5&0.5&&&0.96 &0.54 &0.36&1.51\\
               Figure \ref{cmcutoff-20-19}&&64&0.01&63&1&sigmoid&&&&-2.00&-1.9&&&0.29&0.91&0.91&1.00\\
                 Figure \ref{largecutoff}&&32&0.09 &854 &3 &ReLU &&&&-2.0&2.0&&&0.31 &0.92 &0.92 & 1.0\\
                 Figure \ref{cmcutoff-0507}&&64&0.07 &361 &1 &sigmoid&&&&-0.5 &0.7 &&&1.15 &0.61 &0.48 & 1.28\\
                                Figure \ref{large_mac_ratio}&&128&0.01 &604 &1 &sigmoid &&&&-0.4 &0.4 &&&0.84 &0.60 &0.37 &1.61\\
	\end{tabularx}
        \caption{Configuration and results for case study models. {\bf mb} denotes minibatch size; {\bf lr} denotes learning rate; {\bf hs} denotes hidden layer size; {\bf nl} denotes number of layers; {\bf ha} denotes hidden activation function; {\bf L} and {\bf U} denote lower and upper cutoff ranges; {\bf loss} denotes average test set cross-entropy between true and predicted values; {\bf acc} denotes accuracy; {\bf macc} denotes majority class accuracy; {\bf ratio} is acc/macc.}
%    accuracy, \textbf{macc}: majority class accuracy and
%    \textbf{ratio}: $\frac{\textbf{acc}}{\textbf{macc}}$.} 
\label{tab:perf}
\end{table*}
  \begin{figure}
    \includegraphics[width=0.75\textwidth]{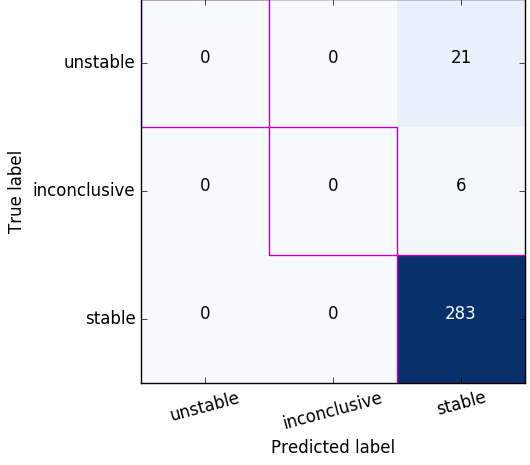}
    \caption{Test set confusion matrix for cutoffs -2.0, -1.9.}
    \label{cmcutoff-20-19}
  \end{figure}    

\section{Results and Discussion}

\begin{figure}[H]
  \centering
    \includegraphics[width=0.75\textwidth]{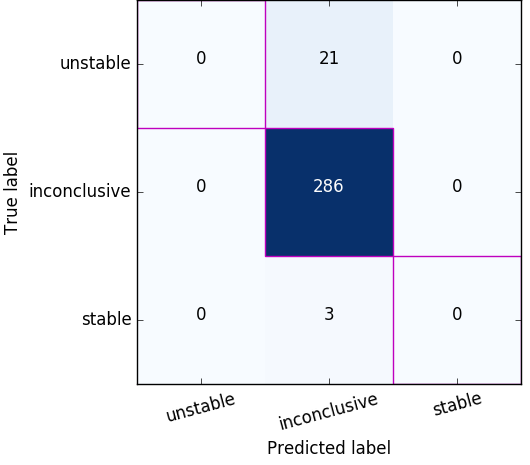}
    \caption{Test set confusion matrix with an unrealistic inconclusive range (-2.0,2.0) where most mutations are labeled as inconclusive.}
    \label{largecutoff}
\end{figure}

\begin{figure}
  \centering
    \includegraphics[width=0.75\textwidth]{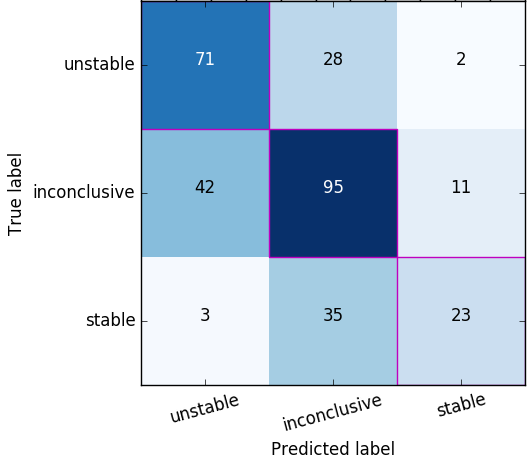}
    \caption{Test set confusion matrix for cutoffs -0.5 and 0.7, where false
      positive and false negative errors (top-right and bottom-left, respectively)
      are minimized.}
    \label{cmcutoff-0507}
\end{figure}

Table \ref{tab:perf} reports hyper-parameters as well as several performance metrics for our models.
The confusion matrices shown in Figures \ref{cmcutoff0505}--\ref{cmcutoff-0507} further elucidate these models' performances.

We first note that when the vast majority of the $\Delta\Delta G$ values fall within a single
region determined by the cutoff boundaries, a classification model can trivially achieve high
accuracy by learning to predict the majority class.  However, labels thus determined
may be impractical for scientific pursuits.  These situations are characterized by 
a high proportion of data points which fall into the majority class giving a high majority class accuracy (\texttt{macc}), 
which is indicated in Table \ref{tab:perf}. 
One such example is given in Figure \ref{cmcutoff-20-19}
which has a small range of indeterminacy, $ [-2, -1.9]$, with a large negative offset. For these bounds, $\texttt{macc} = 91\%$,
 with only 6 inconclusive examples and 21 destabilizing examples. We can see from the confusion matrix that all examples 
 were predicted as stabilizing mutations giving a 91\% accuracy which amounts to a clearly unhelpful classifier. 
Another example of ill-conditioned labeling is shown in Figure     \ref{largecutoff}. 
In this case the indeterminate range is ostensibly too large, $[-2, 2]$ as the model has learned to classify most examples as 
inconclusive.

 Figure
\ref{cmcutoff0505} shows performance for ternary
classification using the traditional $\Delta\Delta G$ range for exclusion of examples, $[-0.5, 0.5]$. If we exclude the somewhat innocuous mistakes of  examples which are incorrectly classified as inconclusive, along with the examples labeled as inconclusive which would be excluded in the traditional approach in the first place, and attend only to egregious mis-classification of stabilizing as non-stabilizing and vice-versa we achieve a 92.2\% accuracy.
From this method of preference, running counter to common practice, the optimal ranges for excluding $\Delta\Delta G$ are not necessarily
centered on zero. 

\begin{figure}
  \centering
    \includegraphics[width=0.75\textwidth]{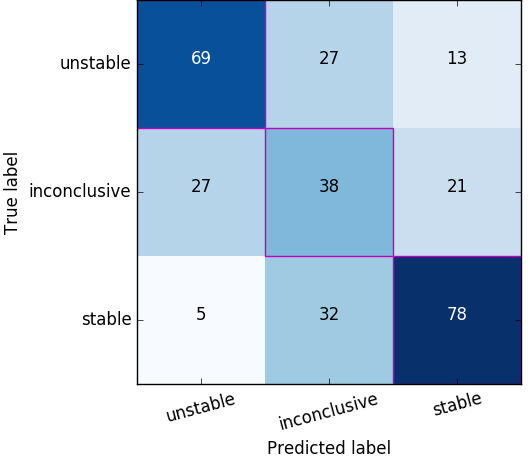}
    \caption{Test set confusion matrix for cutoffs -0.4 and 0.4, where the ratio of accuracy to majority class is maximized.}
    \label{large_mac_ratio}
\end{figure}

For instance, based on this criterion of binary predictions within the ternary classification schema, 
the best cutoff classification range from our experiments is shown in Figure \ref{cmcutoff-0507} with an 
inconclusive range $ [-0.5, 0.7]$, giving a 94.4\% accuracy for the 
binary subset classification task.  
On the same test set, for the ternary task, that model achieved an accuracy of
61\%. Upon initial assessment this performance does not seem great on its own, but we are more concerned with
the model's classification of a destabilizing mutation as a
stabilizing one, and vice versa, than we are of it mis-classifying an
inconclusive mutation. In this case we see that
for this cutoff range the model yields impressive
mis-classification rates of 2\% for destabilizing to stabilizing and 4\%
 for stabilizing to destabilizing. Such low rates of mis-classification
across the inconclusive zone help motivate these findings and suggest
that this range is a potentially good $\Delta\Delta G$ cutoff set.

On the other hand, another promising criterion for optimal cutoff is be the ratio of accuracy (\texttt{acc}) to majority class accuracy, $\texttt{ratio} = \frac{\texttt{acc}}{\texttt{macc}}$, also displayed in Table~\ref{tab:perf}. For any acceptable model this value should be greater than 1, with larger values being better. Figure \ref{large_mac_ratio} shows performance for a model with inconclusive range $[-0.4, 0.4]$ and a significantly higher \texttt{ratio} value than the traditional cutoff

\section{Conclusion and Future Work}
As an extension on our prior work we were interested in assessing the
potential of a deep neural network for classifying the effects of mutations.
We performed a systematic search of the $\Delta\Delta G$
classification cutoff ranges in order to assess the potential
viability of a deep neural network ternary classification approach to
predicting of mutation affects. Rather than simply accept the general
heuristic for classification boundaries of stabilizing, destabilizing
or inconclusive, we strove for a more systematic approach. While our
findings suggest that the heuristic of -0.5 to 0.5 is not a poor
choice by any means, we proposed some compelling arguments for choosing 
other ranges as boundary conditions for $\Delta\Delta G$ values, namely it is most important to
minimize false positive and false negative rates on the ternary task, and maximizing the 
ratio of accuracy to majority class accuracy are both more important metrics to consider besides
accuracy.  

%did find some additional interesting ranges  so that a model
%trained on such a three class label system would have a good accuracy
%while minimizing false positives and false negatives.

%Through the use confusion matrices we assessed several good candidates
%for good $\Delta\Delta G$ classification cutoff ranges. We also
%revealed several ranges of labels that would not be particularly
%useful scientifically though they produce high model accuracy.

For future work we plan to develop robust algorithmic approaches to
assess the likely cutoff ranges in ML-based models. We are currently
in the development of an end-to-end differentiable approach to jointly
learn an optimal cutoff range alongside DNN parameters, as opposed to
relying on a parameter sweep as in the current work. Also, expanding
our data set with additional mutation $\Delta\Delta$G data -- data for
proteins with multiple mutations -- will likely enhance the DNN's
learning and ultimately increase accuracy. We also hope to expand our
study into other machine learning algorithms.

\subsection*{Acknowledgments}
The authors would like to thank the Nvidia corporation for donating a Titan Xp GPU used in this research.

{\small
  \balance
\bibliographystyle{plain}
\bibliography{bio.bib}
}

\end{document}